\documentstyle[preprint,aps,psfig]{revtex}
\begin{document}
\tighten
%
% uncomment \draft to have PACS numbers appear
%\draft
% put preprint number(s).
%\preprint{}
\title{$\Lambda-\Sigma^0$ mixing in finite nuclei}
\author{H.~M\"uller and J.~R.~Shepard}
\address{Department of Physics, University of Colorado,
Boulder, Colorado 80309}
\vskip1in
%\date{\bf DRAFT: \today}
\date{\today}
\maketitle
\begin{abstract}
{Properties of hypernuclei are studied in the context of a chiral
Lagrangian which successfully describes ordinary nuclei.
$\Lambda-\Sigma^0$ mixing arises from nondiagonal vertices in
flavor space induced by the vector mesons and by the electromagnetic
field.
The set of Dirac equations for the coupled hyperon system is discussed.
Results are presented for energy spectra and electromagnetic properties of 
hyperons in the nuclear environment. 
Simple estimates suggest that flavor mixing can lead to 
sizable changes in the lifetimes of $\Lambda$ hypernuclei.}
\end{abstract}
\vspace{20pt}
\pacs{PACS number(s): 21.80.+a}
%
%###########################################################################%
%###########################################################################%
%
\section{Introduction}
Hyperons embedded in the nuclear medium provide an unique tool 
for studying
the elementary nucleon-hyperon and hyperon-hyperon interaction.
On the experimental side, physics of hypernuclei has entered a phase in which
not only ground state energies but also excitation spectra and decay
properties are being measured \cite{BANDO90}.
Theoretically, binding energies and single particle spectra of
hypernuclei are studied in a variety of nonrelativistic \cite{RAYET81} and
relativistic \cite{RUFA87,MARES89} mean-field models. 
The relativistic models usually involve
the baryon octet and several strange
\cite{SCHAFFNER93,PAPAZOGLOU98a,PAPAZOGLOU98b,MUELLER99} and nonstrange mesons.

The concepts and methods of effective field theory (EFT) 
have lead to new insights into relativistic mean field models for nuclear
structure \cite{FST97,SEROT97}.
For ordinary nuclei a chiral effective Lagrangian has been constructed 
which successfully reproduces low-energy nuclear phenomenology \cite{FST97}.
One important implication is that all interaction terms that are consistent 
with the underlying symmetries of QCD should be included. 

Recently, this approach was generalized and applied to strange nuclear
matter \cite{MUELLER99}.
It was demonstrated that a framework which includes the most general types 
of interactions leads to new and interesting many-body effects. 
Most prominently, D-type couplings between baryons and 
vector mesons give rise to $\Lambda-\Sigma^0$ flavor mixing. 
In the present work we extend the analysis to finite systems
and include various tensor couplings and couplings to the electromagnetic 
field which are relevant in hypernuclei.
Following Ref.~\cite{FST97}, the low-energy electromagnetic structure of 
the hyperons is described within the effective Lagrangian approach
so that no external form factors are needed. 

The central point in the present discussion is the 
analysis of $\Lambda-\Sigma^0$ flavor mixing.
The primary result is that a hypernucleus is generally in a state of mixed 
flavor rather than in a state with distinct $\Lambda$ and $\Sigma^0$ particles. 
A particle interpretation is only possible in terms of the actual energy
eigenstates which are a superposition of the flavor eigenstates.
As a consequence, systems which contain $\Lambda$ hyperons always have
a small admixture of $\Sigma^0$ hyperons, and vice versa.
The physical nature of this effect is similar to the recently much discussed 
neutrino oscillations \cite{BILENKY78}.

Flavor mixing is mainly driven by the mean field of the $\rho$ meson and 
signatures of the effect are more likely to be observed in very heavy 
asymmetric nuclei.
We therefore focus on hypernuclei that consist of a $^{208}Pb$ core 
and examine single particle spectra, electromagnetic properties and lifetimes.
The basic result of our analysis is that the flavor mixing is a 
rather small effect. 
The shifts in the single-particle
spectra are well below typical hypernuclear spin-orbit splittings.
A more direct signature of the mixing are deviations of hypernuclear
magnetic moments from their single-particle values. Our results indicate
changes at the $2\%$ level.
Because of the hyperon mass differences, $\Lambda$ and $\Sigma^0$ hypernuclei
are characterized by strikingly different lifetimes and widths which
can be as much as twelve orders of magnitude apart.
Although the admixture of a $\Sigma^0$ in a $\Lambda$ hypernucleus is
very small, the dramatic difference in the decay width leads to sizable
effects. Based on simple assumptions we show that the width of 
the $\Lambda$ in excited states can be significantly enhanced
by flavor mixing.

The outline of this paper is as follows:
In Sec.~\ref{lagrangian}, we present the relevant part of the
effective Lagrangian. 
Section \ref{mixing} contains a discussion of the Dirac equation which
governs the coupled hyperon system.
In Sec.~\ref{results} we discuss the impact of the flavor mixing
on specific properties of hypernuclei.
Section~\ref{summary} contains a short summary. 
%
%###########################################################################%
%###########################################################################%
%
\section{The effective Interaction}
\label{lagrangian}
The model is based on an effective Lagrangian that realizes chiral
symmetry and vector meson dominance.
In the nucleon sector this approach has been successful in describing the
properties of ordinary nuclei \cite{FST97,SEROT97}.
More recently, these ideas have been generalized to describe strange nuclear
matter \cite{MUELLER99}.
In our contribution we will extend this analysis and develop new aspects
which arise in strange finite systems.
Most of the material can be found in Ref.~\cite{MUELLER99} and we will
quote only the new ingredients relevant for hypernuclei.

The effective degrees of freedom are the baryon octet and the
vector meson nonet.
The baryons and the vector meson octet are collected in $3\times3$ 
traceless hermitian matrices
\begin{equation}
B = \left(\begin{array}{ccc}
\frac{1}{\sqrt{6}}\Lambda+\frac{1}{\sqrt{2}}\Sigma^0 & \Sigma^+  & p \\
\Sigma^- & \frac{1}{\sqrt{6}}\Lambda-\frac{1}{\sqrt{2}}\Sigma^0 &  n \\
\Xi^- & \Xi^0& -\frac{2}{\sqrt{6}}\Lambda\\
\end{array} \right) \ ,
\label{eq:baryon}
\end{equation}
\begin{equation}
V_\mu = \left(\begin{array}{ccc}
\frac{1}{\sqrt{6}}V^8_\mu+\frac{1}{\sqrt{2}}\rho^0_\mu & \rho^+_\mu 
& K^{*+}_\mu \\
\rho^-_\mu & \frac{1}{\sqrt{6}}V^8_\mu-\frac{1}{\sqrt{2}}\rho^0_\mu 
&  K^{*0}_\mu \\
K^{*-}_\mu & \overline{K}^{*0}_\mu& -\frac{2}{\sqrt{6}}V^8_\mu\\
\end{array} \right) \ .
\label{eq:vector}
\end{equation}
The physical $\omega$ and $\phi$ mesons arise 
from the mixing of the $V^8_\mu$ and the vector meson singlet $S_\mu$ via
\begin{eqnarray}
\omega_{\mu}&=& \cos(\theta) S_{\mu} + \sin(\theta) V^8_{\mu}  \ ,
\label{eq:mesonmix} \\
\phi_{\mu}&=& \sin(\theta) S_{\mu} - \cos(\theta) V^8_{\mu}  \ .
\nonumber
\end{eqnarray}
We also include a light isoscalar scalar meson $\varphi$ 
which simulates the exchange of correlated pions and kaons.

The couplings of the vector mesons to the baryons are contained in
\begin{eqnarray}
{\cal L}_{BMB} &=& 
- g_F {\rm Tr}\Bigl({\overline B}[V\!\!\!\!/, B]\Bigr)
        - g_D {\rm Tr}\Bigl({\overline B}\{V\!\!\!\!/, B\}\Bigr)
        - g_S {\rm Tr}\Bigl({\overline B}S\!\!\!/B\Bigr) 
\label{eq:lbvb}\\
&&\null - {f_F\over 4M} {\rm Tr}\Bigl({\overline B}
          [\sigma_{\mu\nu}V\!\!\!\!/^{\mu\nu}, B]\Bigr)
        - {f_D \over 4M}{\rm Tr}\Bigl({\overline B}
          \{\sigma_{\mu\nu}V\!\!\!\!/^{\mu\nu}, B\}\Bigr)
        - {f_S\over 4M} {\rm Tr}\Bigl({\overline B}
          \sigma_{\mu\nu}S\!\!\!\!/^{\mu\nu} B\Bigr) \ .
\nonumber
\end{eqnarray}
Couplings to the electromagnetic field are introduced by
\begin{eqnarray}
{\cal L}_{EM} &=& 
- e {\rm Tr}\Bigl({\overline B}[Q A\!\!\!\!/,B]\Bigr)
- e{\mu_D\over 4M} {\rm Tr}\Bigl({\overline B}\{{\cal Q} A\!\!\!\!/, B\}\Bigr)
- e{\mu_F\over 4M} {\rm Tr}\Bigl({\overline B}[{\cal Q} A\!\!\!\!/, B]\Bigr)
\label{eq:lem}\\
&&\null 
+ e{\beta_D\over 2M^2} {\rm Tr}\Bigl({\overline B}\gamma^\nu\partial^\mu
F_{\mu\nu} \{{\cal Q} , B\}\Bigr)
+ e{\beta_F\over 2M^2} {\rm Tr}\Bigl({\overline B}\gamma^\nu\partial^\mu
F_{\mu\nu} [{\cal Q} , B]\Bigr) \ ,
\nonumber
\end{eqnarray}
where ${\cal Q}={\rm diag}\{2/3,-1/3,-1/3\}$ is the quark charge matrix.
Combined with vector meson dominance, the Lagrangian Eq.~(\ref{eq:lem}) 
describes the low energy electromagnetic structure of the baryons
so that external form factors are not needed.
For example, the tensor couplings generate the magnetic moments of the baryons.
In the vacuum Eq.~(\ref{eq:lem}) implies the Coleman and Glashow
\cite{COLEMAN61} relations,
\begin{equation}
{1\over 2}\mu_{n}=\mu_{\Lambda}=-\mu_{\Sigma^0}=
-{1\over\sqrt{3}}\mu_{\Lambda\Sigma^0}  \ .
\label{eq:cc}
\end{equation}
To constrain the couplings we follow closely Ref.~\cite{MUELLER99}.
For the meson-baryon couplings we assume $SU(3)$ symmetry and that the OZI rule
holds, {\em i.e.} the couplings between nucleons and the $\phi$ meson
vanish. 
Furthermore, relation Eq.~(\ref{eq:mesonmix}) is implemented with
the ideal mixing angle $\sin(\theta)=1/\sqrt{3}$.
%We assume ideal mixing in the relation Eq.~(\ref{eq:mesonmix}).
For given values of the corresponding $\omega$ and $\rho$ 
couplings to the nucleons, the set of parameters 
$(g_F, g_D, g_S, f_F, f_D, f_S)$ is then fixed ($M$ is taken to be the nucleon
mass).

The Coleman and Glashow relations Eq.~(\ref{eq:cc}) 
hold only in the strict $SU(3)$ limit.
The physical values of the magnetic moments can be generated 
%In reality there are considerable deviations which can be taken into account 
by adding appropriate symmetry breaking terms.
We use the Particle Data Group \cite{PDG} values
\begin{equation}
\mu_{\Lambda}=-0.613 \quad , \quad
\mu_{\Lambda\Sigma^0}=1.61   \ .
\nonumber
\end{equation}
For the magnetic moment for the $\Sigma^0$, 
which is experimentally not accessible, we employ the result
of the chiral perturbation theory calculation in Ref.~\cite{MEISSNER97}
\begin{equation}
\mu_{\Sigma^0}=0.65  \ .
\nonumber
\end{equation}
The parameters $\beta_F$ and $\beta_D$ contribute to the charge radii of 
the baryons which are not known in the hyperon sector.
For simplicity, we assume $SU(3)$ symmetry 
which determines the parameters from the corresponding values in the 
nucleon sector.
To specify the coupling of the $\Lambda$ to the scalar field $\varphi$ 
we fit our model to reproduce the lowest $\Lambda$ level in
$^{17}_{\Lambda}O$ which can be extracted from the experimental results
in Ref.~\cite{CHRIEN88}.
For the $\Sigma^0$ we follow the 
phenomenological approach of Refs.~\cite{MUELLER99,SCHAFFNER94} and
require that the coupling reproduces the hyperon potential 
in nuclear matter which is taken to be
\begin{equation}
U^{\Sigma}= g^\varphi_\Sigma \varphi - g^\omega_\Sigma \omega^0 
= 25 {\rm MeV} \ .
\nonumber
\end{equation}
In the nucleon sector we employ the parameter set G1 of Ref.~\cite{FST97}.
The corresponding hyperon couplings are listed in Table~\ref{tab:coup}.
%
%###########################################################################%
%###########################################################################%
%
\section{$\Lambda-\Sigma^0$ mixing in finite nuclei}
\label{mixing}
We consider $\Lambda,\Sigma^0$ hypernuclei in a mean field approximation
based on the interaction terms introduced in the last section.
We are primarily interested in studying the new features arising from the
flavor mixing. 
As a simple approximation we neglect the 
response of the nuclear core to the hyperons so that the hyperons can 
be treated separately. As the first step in this valence-hyperon approximation
we perform the calculation 
for the nuclear core and substitute the resulting meson mean fields in the
Dirac equations that govern the hyperons.
To maintain rotational invariance we consider doubly magic nuclei. 

The nondiagonal terms in flavor space given in Eq.~(\ref{eq:lbvb}) and 
Eq.~(\ref{eq:lem}) mix the $\Lambda$ and $\Sigma^0$. 
The wave function of the combined system can be 
written as
\begin{equation}
\Psi ({\bf r},t) = e^{-iEt}\left(\begin{array}{c}
\Psi_\Lambda({\bf r})\\
\Psi_\Sigma({\bf r})\\
\end{array} \right) \ ,
\label{eq:wf0}
\end{equation}
with
\begin{equation}
\Psi_{\Lambda,\Sigma}({\bf r}) = \left(\begin{array}{c}
iG^{\Lambda,\Sigma}_{n\kappa}(r)/r {\cal Y}_{\kappa m}\\
-F^{\Lambda,\Sigma}_{n\kappa}(r)/r {\cal Y}_{-\kappa m}\\
\end{array} \right) \ ,
\label{eq:wfls}
\end{equation}
where $n$ is the principal quantum number and ${\cal Y}_{\kappa m}$ is
a spin$-1/2$ spherical harmonic. The nonzero integer $\kappa$ determines
$j$ and $l$ through $\kappa=(2j+1)(l-j)$.
The separation leads to a coupled set of equations for the radial
wave functions:
\begin{eqnarray}
\Delta_\Lambda^- G^\Lambda 
+ \left[ {d\over dr} - {\kappa\over r} + T_\Lambda \right] F^\Lambda 
&=& V_{\Lambda\Sigma} G^\Sigma - T_{\Lambda\Sigma} F^\Sigma
\label{eq:radial} \\
\Delta_\Lambda^+ F^\Lambda 
- \left[ {d\over dr} + {\kappa\over r} -T_\Lambda \right] G^\Lambda 
&=& V_{\Lambda\Sigma} F^\Sigma -T_{\Lambda\Sigma} G^\Sigma \\
\Delta_\Sigma^- G^\Sigma 
+ \left[ {d\over dr} - {\kappa\over r} + T_\Sigma \right] F^\Sigma 
&=& V_{\Lambda\Sigma} G^\Lambda - T_{\Lambda\Sigma} F^\Lambda\\
\Delta_\Sigma^+ F^\Sigma 
- \left[ {d\over dr} + {\kappa\over r} -T_\Sigma \right] G^\Sigma 
&=& V_{\Lambda\Sigma} F^\Lambda - T_{\Lambda\Sigma} G^\Lambda
\end{eqnarray}
where we have introduced the notation
\begin{equation}
\Delta_{\Lambda,\Sigma}^\pm=E-V_{\Lambda,\Sigma}\pm(M_{\Lambda,\Sigma}
-g_{\Lambda,\Sigma}^\varphi \varphi) \ .
\label{eq:delta}
\end{equation}
The individual vector and tensor potentials are listed in Table~\ref{tab:pots}.
Bound state solutions are normalized according to
\begin{equation}
\int_0^\infty dr \bigr(|G^\Lambda|^2+|F^\Lambda|^2 \bigl)
+\int_0^\infty dr \bigr(|G^\Sigma|^2+|F^\Sigma|^2\bigr) \equiv
N_\Lambda + N_\Sigma = 1  \ ,
\label{eq:norm}
\end{equation}
so that the system contains one strange baryon.
For bound states we seek solutions that are regular at the origin and that 
fall off sufficiently fast at infinity. The behavior at the origin can be
studied by assuming constant values of the vector and scalar potentials 
and zero values for the tensor terms. In this case, the set of equations 
(\ref{eq:radial}) is easily solved by writing
\begin{equation}
G^{\Lambda,\Sigma}= g^{\Lambda,\Sigma}r j_l (\alpha r)
\quad , \quad
F^{\Lambda,\Sigma}= f^{\Lambda,\Sigma}r j_{l\pm1}(\alpha r)
\label{eq:ansatz} \ ,
\end{equation}
with $l+1$ for $\kappa<0$ and $l-1$ for $\kappa>0$.
Substituting Eq.~(\ref{eq:ansatz}) into Eq.~(\ref{eq:radial}) leads to a set of
algebraic equations for the unknown coefficients $g^{\Lambda,\Sigma}$ and
$f^{\Lambda,\Sigma}$. Solutions exist only if the condition
\begin{equation}
\alpha^4-\alpha^2 (\Delta_{\Lambda}^+ \Delta_{\Lambda}^-
+\Delta_{\Sigma}^+ \Delta_{\Sigma}^- + 2 V_{\Lambda\Sigma}^2)
+(\Delta_{\Lambda}^- \Delta_{\Sigma}^- - V_{\Lambda\Sigma}^2)
(\Delta_{\Lambda}^+ \Delta_{\Sigma}^+ - V_{\Lambda\Sigma}^2)=0
\label{eq:alpha0} \ ,
\end{equation}
is fulfilled. The roots of this equation constitute two qualitatively
different solutions. A type of solutions
which reduces to a pure $\Lambda$ with
$g^\Sigma=f^\Sigma=0$ if the mixing is turned off,
and a second type of solutions which reduces to a pure $\Sigma^0$
with $g^\Lambda=f^\Lambda=0$.
A general solution to the radial equations consists of a super position
of the two types; the relative weight has to be determined numerically.

The radial equations decouple in the asymptotic regime $r\to \infty$
and, because there is no direct electromagnetic coupling,
can be replaced by the corresponding free equations. The solutions
are
\begin{equation}
G^{\Lambda,\Sigma}\mathop{=}_{r\to \infty} 
c^{\Lambda,\Sigma} \sqrt{r} K_{l+1/2} (\beta^{\Lambda,\Sigma}r)
\quad , \quad
F^{\Lambda,\Sigma}\mathop{=}_{r\to \infty} 
-c^{\Lambda,\Sigma} 
\left[ r {M_{\Lambda,\Sigma}-E \over M_{\Lambda,\Sigma} + E}\right]^{1/2}
K_{l\pm 1+1/2} (\beta^{\Lambda,\Sigma}r) \ ,
\label{eq:asymp}
\end{equation}
with $\beta^{\Lambda,\Sigma}=\sqrt{M_{\Lambda,\Sigma}^2 - E^2}$.
The asymptotic coefficients $c^{\Lambda}$ and $c^{\Sigma}$ together
with the relative weight at the origin and the energy variable $E$ 
constitute four
unknowns which may be determined by matching the two large and two small radial 
wave functions at some matching radius.

The energy spectrum depends on the strength of the mixing potentials
and on the mass difference of the two hyperons. 
When mixing is suitably weak as here, there are two
qualitative different sets of solutions. Solutions dominated by the 
$\Lambda$ part
of the wave function ($\Lambda$-like) that can be characterized by
\begin{equation}
N_\Lambda \gg N_\Sigma
\label{eq:norml} \ ,
\end{equation}
and $\Sigma$-like solutions with
\begin{equation}
N_\Sigma \gg N_\Lambda
\label{eq:norms} \ .
\end{equation}
Because the energy eigenstates contain both flavors, a hypernucleus is 
generally in a state of mixed flavor rather
than in a state with distinct $\Lambda$ and $\Sigma^0$ particles.
A (quasi) particle interpretation is only possible in terms of the actual 
energy eigenstates,  $\Lambda$-like or $\Sigma$-like, which are not
flavor eigenstates.
A (time-dependent) flavor eigenstate can only arise as a superposition
of energy eigenstates \cite{MUELLER99}.

$\Lambda-\Sigma^0$ mixing is superficially similar to the phenomenon of
neutrino flavor mixing which gives rise to neutrino oscillations.
However, the origin of both effects is fundamentally different.
Neutrino oscillations are assumed to occur in the vacuum arising
from a nondiagonal mass matrix in flavor space which contains the vacuum mass
parameters \cite{BILENKY78}. This effect can be appreciably enhanced when
neutrinos pass through dense matter as predicted by the MSW effect \cite{MSW}.
In contrast, the $\Lambda-\Sigma^0$ mixing is a {\em true many body effect}
arising from the nondiagonal self energies which are generated by the nuclear 
medium. As long as small isospin violations can be neglected,
the vacuum self energies are diagonal in flavor space and the asymptotic
states can be properly identified as the pure $\Lambda$ and $\Sigma^0$
flavor states.
%
%###########################################################################%
%###########################################################################%
%
\section{Results and Discussion}
\label{results}
\subsection{Ground state Properties}
Armed with an understanding of the general features of the solutions,
we now turn to specific properties of hypernuclei.
Flavor mixing is primarily driven by the mean field of the neutral
$\rho$ meson and the effect is very small in light and symmetric
nuclei. To generate the mean fields we therefore 
consider a $^{208}Pb$ core nucleus.

Strict $\Sigma$-like bound states only exist if the mass
difference between the two hyperons is small. 
In the chiral limit with equal $\Lambda$ and $\Sigma^0$ masses
two series of bound states, $\Lambda$-like and
$\Sigma$-like, arise. 
Increasing the mass of the $\Sigma^0$ increases
the energy of the states in the $\Sigma$-like part of the spectrum.
At some point the energy of the most weakly bound $\Sigma$-like state 
exceeds the mass of the $\Lambda$, whereupon the parameter 
$\beta^{\Lambda}=\sqrt{M_{\Lambda}^2 - E^2}$ in Eq.~(\ref{eq:asymp}) turns
imaginary and the $\Lambda$ part of the wave function has to 
be replaced by a continuum wave function.
Using the physical values for the hyperon masses, we find that, for
all $\Sigma$-like states, the $\Lambda$ part of the wave function is
unbound. As a consequence, only true bound states of $\Lambda$-like
configurations exist. 

The impact of the flavor mixing on the energy spectrum is rather small.
The situation is illustrated in Fig.~\ref{fig:spec}. To make it easier to
study the effect we have scaled the mixing potentials
$T_{\Lambda\Sigma}$ and $V_{\Lambda\Sigma}$ by a factor $\zeta$.
Part (a) indicates the energy of the lowest $\Lambda$ like states.
The mixing is attractive and and tends to lower the binding energies.
However, for the realistic situation $\zeta=1$ the effect is very small, 
the energy levels decrease by less than $ 0.01$ MeV.
To set the proper scale this number has to be compared to typical
hypernuclear spin-orbit splittings 
$(\Delta E{\lower0.6ex\vbox{\hbox{$\ \buildrel{\textstyle <}
         \over{\sim}\ $}}} 0.5)$MeV
that can be resolved experimentally.
The parabolic shape of the curves reflects the fact 
that the mixing is a second order effect in perturbation theory.
For the $\Lambda$ the tensor couplings significantly reduce the 
spin-orbit potential which leads to the very small splittings of the 
$(1p_{3/2},1p_{1/2})$ and $(1d_{5/2},1d_{3/2})$ for zero and weak mixing.
Part (b) of Fig.~\ref{fig:spec} indicates the values of
$N_\Lambda$ and $N_\Sigma$ as defined in Eq.~(\ref{eq:norm})
for the lowest $s_{1/2}$ state. Other states show very similar
behavior. For the realistic situation $\zeta=1$ the system
consists primarily of a $\Lambda$ with a very small admixture ($ 0.003\%$)
of the $\Sigma^0$ flavor.

A direct indication of the flavor mixing is a nonvanishing mixed 
baryon density
\begin{equation}
\rho_{\Lambda\Sigma}= < \bar{\Psi}_\Lambda \gamma^0 \Psi_\Sigma>
+ < \bar{\Psi}_\Sigma \gamma^0 \Psi_\Lambda> \ ,
\label{eq:rhols}
\end{equation}
which is shown in Fig.~\ref{fig:rho} together with the pure flavor
densities
\begin{equation}
\rho_{\Lambda,\Sigma}= < \bar{\Psi}_{\Lambda,\Sigma}
\gamma^0 \Psi_{\Lambda,\Sigma}> \ .
\label{eq:rho}
\end{equation}
Although only $1\%$ of the size of the $\Lambda$ density,
the mixed density is much bigger than the $\Sigma$ density.
This is because $\rho_{\Lambda\Sigma}$ arises from the amplitude 
for a $\Sigma^0$ admixture whereas $\rho_{\Sigma}$ arises from the 
$\Sigma^0$ probability.

We now turn to electromagnetic properties. In the $\Lambda\Sigma^0$ sector
the electromagnetic current obtained from Eq.~(\ref{eq:lem}) is given
by 
\begin{eqnarray}
J^\mu_{EM}&=&-{\mu_\Lambda \over 2M} \partial_\nu \Bigl (
\bar{\Psi}_\Lambda \sigma^{\nu\mu} \Psi_\Lambda \Bigr)
-{\mu_\Sigma \over 2M} \partial_\nu \Bigl (
\bar{\Psi}_\Sigma \sigma^{\nu\mu} \Psi_\Sigma  \Bigr)
-{\mu_{\Lambda\Sigma} \over 2 M} \partial_\nu \Bigl (
\bar{\Psi}_\Lambda \sigma^{\nu\mu} \Psi_\Sigma
+\bar{\Psi}_\Sigma \sigma^{\nu\mu} \Psi_\Lambda \Bigr)
\label{eq:jem} \\
&&\null +
{\beta_D \over 6M^2} \partial^2 \Bigl (
\bar{\Psi}_\Lambda \gamma^\mu \Psi_\Lambda
-\bar{\Psi}_\Sigma \gamma^\mu \Psi_\Sigma  \Bigr)
-{\beta_D \over 2 \sqrt{3} M^2} \partial^2 \Bigl (
\bar{\Psi}_\Lambda \gamma^\mu \Psi_\Sigma
+\bar{\Psi}_\Sigma \gamma^\mu \Psi_\Lambda \Bigr) \ ,
\nonumber
\end{eqnarray}
where we have disregarded terms that do not contribute in the mean field
approximation.
The corresponding charge density can be decomposed according to the flavor
indices. This is illustrated in Fig.~\ref{fig:rhoc}. Indicated are 
two states with total angular momentum $j=1/2$ for which the charge 
density is radially symmetric.  Part (a) shows the
various contributions for a $1s_{1/2}$ state. 
The charge density is dominated by the $\Lambda$; the contribution of the
$\Sigma^0$ is again rather small. 
The mixed density leads to a sizable increase near the origin.
Part (b) shows the same situation for the $1p_{1/2}$ state.
Here the mixed density contribution is less important.

We can determine the magnetic moments of the hypernucleus from the current
in Eq.~(\ref{eq:jem}). The Coleman-Glashow relations in Eq.~(\ref{eq:cc}) 
suggest that the mixing leads to deviations of the moments from the
pure flavor values \cite{DOVER95}.
For the magnetic moments, the mixing induces changes which are of first
order. The impact is therefore much greater than for the energy levels.
This can be studied in Fig.~\ref{fig:mag} which indicates
moments for different levels. Similar as in Fig.~\ref{fig:spec} we have 
scaled the mixing potentials by a factor $\zeta$. 
Changes with respect to the single particle value are mainly induced by the
transition moment $\mu_{\Lambda\Sigma}$ 
decreasing the size of the moments for the $\Lambda$-like states.
For very large mixing the moments even change sign.
However, in the relevant region
$(\zeta=1)$ the deviations of the moments from their pure flavor values
are only $ \approx2\%$. 
In our simple valence-hyperon approximation
the magnetic moments are very close to the corresponding
nonrelativistic Schmidt values.
As demonstrated in \cite{GATTONE91,COHEN92} this remains true when
the response of the nuclear core is taken into account. This is in
contrast to the nucleon case where the core response has to be included
properly to restore the magnetic moments to their Schmidt values
\cite{SHEPARD88}.

Although there has been no measurement of hypernuclear magnetic moments, 
proposals have been made  for future experiments \cite{NAKAI96}.
Our results indicate that very precise measurements are necessary in order
to find signatures of the flavor mixing.
For comparison, the experimental uncertainties of $\Lambda$ magnetic moments 
in free space are of the order
$\Delta\mu_\Lambda{\lower0.6ex\vbox{\hbox{$\ \buildrel{\textstyle <}
         \over{\sim}\ $}}} 1\%$ \cite{PDG}.
%
%###########################################################################%
%###########################################################################%
%
\subsection{Continuum Solutions}
As mentioned earlier, there are no bound $\Sigma$-like solutions
for the physical values of the hyperon masses.
However, signatures of
bound pure $\Sigma^0$ states, which arise when the mixing is turned off, 
can be found as resonances.
In the energy range $M_\Lambda \leq E \leq M_\Sigma$, the asymptotic 
form for the $\Lambda$ part of the wave function in Eq.~(\ref{eq:asymp})
has to be replaced by
\begin{equation}
\left(\begin{array}{c}
G^\Lambda\\
F^\Lambda\\
\end{array} \right)
\mathop{=}_{r\to \infty}\alpha^\kappa_j r
\left(\begin{array}{c}
j_l(\beta_\Lambda r)\\
\mp \left({E-M_\Lambda\over E+M_\Lambda}\right)^{1/2} 
j_{l\pm1}(\beta_\Lambda r)\\
\end{array} \right)
+\alpha^\kappa_y r
\left(\begin{array}{c}
y_l(\beta_\Lambda r)\\
\mp \left({E-M_\Lambda\over E+M_\Lambda}\right)^{1/2} 
y_{l\pm1}(\beta_\Lambda r)\\
\end{array} \right)  \ ,
\label{eq:wfcl}
\end{equation}
with $\beta_\Lambda=\sqrt{E^2-M_\Lambda{}^2}$.
The numerical coefficients $\alpha^\kappa_{j,y}$ determine the phase
shifts $\delta^\kappa$ and are calculated by matching the wave 
functions similar as 
discussed in Sect.~\ref{mixing} for the bound states.
The function $\sin^2(\delta^\kappa)$ for a $1p_{3/2}$ state is indicated in
Fig.~\ref{fig:sin2}. Two bound states which arise for a
pure $\Sigma^0$ flavor lead to very narrow resonances at 
$E\approx 1174.16$MeV and $E\approx 1184.64$MeV.
The complete hyperon spectrum for the $^{208}Pb$ core nucleus is 
indicated in Fig~\ref{fig:spec2}. On the left-hand side are the truly bound
$\Lambda$-like states and on the right-hand side the resonance energies of the 
$\Sigma$-like configurations.
The figure nicely displays the quite different character of the spin-orbit
force \cite{MARES94} for the two flavors.
For the $\Lambda$-like states it leads to a sizable reduction
of the splittings and an increase for the $\Sigma$-like states.

For energies $E \geq M_\Sigma$ two physical different situations can
be realized. $\Lambda$-like scattering states characterized
by the asymptotic form given in Eq.~(\ref{eq:wfcl}) and by
\begin{equation}
\left(\begin{array}{c}
G^\Sigma\\
F^\Sigma\\
\end{array} \right)
\mathop{=}_{r\to \infty}\gamma^\kappa_y r
\left(\begin{array}{c}
y_l(\beta_\Sigma r)\\
\mp \left({E-M_\Sigma\over E+M_\Sigma}\right)^{1/2} 
y_{l\pm1}(\beta_\Sigma r)\\
\end{array} \right)  \ ,
\label{eq:wfcs}
\end{equation}
for the $\Sigma^0$ part of the wave function.
Eq.~(\ref{eq:wfcl}) together with Eq.~(\ref{eq:wfcs}) describes an incoming
pure $\Lambda$ scattered off the target nucleus. In addition to the $\Lambda$,
the outgoing wave function also contains a (small) contribution of the 
$\Sigma^0$,
indicating a possible transition between a $\Lambda$ and a $\Sigma^0$ during
the collision.
Similarly, $\Sigma$-like scattering states can be described by interchanging the
flavor indices in Eq.~(\ref{eq:wfcl}) and in Eq.~(\ref{eq:wfcs}).

In our simple mean field picture
the $\Lambda$ and $\Sigma^0$ are treated on an equal footing. 
In a more realistic approach various decay mechanisms 
which lead to significantly different properties of $\Lambda$ and $\Sigma$
hypernuclei have to be taken into account.
Hypernuclei are typically produced through hadronic reactions such
as $(K^\pm,\pi^\pm)$. Eventually, they will decay through nonleptonic
weak processes which involve the emission of pions or nucleons.
For $\Lambda$-hypernuclei the dominant process is the nonmesonic
decay mode $\Lambda N\to NN$ which leads to lifetimes comparable to
the lifetime of a free $\Lambda$.
Properties of $\Sigma$-hypernuclei are still controversial.
Complications arise from the strong $\Sigma N\to \Lambda N$ conversion process,
which leads to widths in the excitation spectrum of several MeV.
In a realistic scenario this process completely dwarfs the resonances
indicated in Fig~\ref{fig:sin2} with $\Gamma\approx 1$keV. 
Because of the strikingly different scales set by the $\Lambda$ and
$\Sigma^0$ decay width it is interesting to examine the impact of the
flavor mixing on lifetimes and width of hypernuclei.
Let us assume that the decay of the
$\Lambda$ and $\Sigma^0$ can be described by adding appropriate
optical potentials \cite{DOVER89} to our mean field model. 
The mixing potentials are 
very small and we can use perturbation theory.
To first order a $\Lambda$ like solution can be written as
\begin{equation}
\Psi = \left(\begin{array}{c}
\Psi_\Lambda^0\\
C_{\Lambda\Sigma}\Psi_\Sigma^0\
\end{array} \right) \ ,
\label{eq:pert}
\end{equation}
where the wave functions are normalized pure flavor solutions with
complex energies
\begin{equation}
\bar{E}^0_{\Lambda}=E^0_{\Lambda}-{i\over 2}\Gamma^0_{\Lambda}
\quad , \quad
\bar{E}^0_{\Sigma}=E^0_{\Sigma}-{i\over 2}\Gamma^0_{\Sigma} \ .
\label{eq:ec}
\end{equation}
To second order the width is then given by
\begin{equation}
\Gamma_{\Lambda}=
(1-|C_{\Lambda\Sigma}|^2)\Gamma^0_{\Lambda}+
|C_{\Lambda\Sigma}|^2\Gamma^0_{\Sigma} \ .
\label{eq:shift}
\end{equation}
The coefficient $C_{\Lambda\Sigma}$ which characterizes the 
admixture of the $\Sigma^0$ can be estimated by
\begin{equation}
|C_{\Lambda\Sigma}|^2 \approx {N_\Sigma \over N_\Lambda} \ ,
\nonumber
\end{equation}
with the norms introduced in Eq.~(\ref{eq:norm}).
Using typical numbers \cite{BANDO90} for the in-medium width 
$\Gamma^0_\Sigma\approx 10^{12}\Gamma^0_\Lambda$
this leads to the rough estimate
\begin{equation}
\Gamma_{\Lambda}\approx 10^7 \Gamma^0_\Lambda \ .
\label{eq:gshift}
\end{equation}
Thus, flavor mixing significantly decreases the weak lifetimes of
$\Lambda$-like hypernuclear systems.
The resonances populated in hypernuclear reactions are often highly excited
states which may decay by electromagnetic processes prior to their
weak decays. As argued in Ref.~\cite{DOVER95} the corresponding lifetimes
can also be modified by the flavor mixing. Electromagnetic decays are typically
much slower than the $\Sigma N\to \Lambda N$ conversion. However, they
are of the same order of magnitude for $\Lambda$ and $\Sigma$ sates and,
according to Eq.~(\ref{eq:shift}), flavor mixing can only lead to very small
modifications.

The fact that level mixing can change the lifetime of states
is a well known phenomenon.
For example, isospin mixing in $^{12}C$ has an
appreciable effect on decay rates and form factors
\cite{ADELBERGER77,FLANZ79}.
At this point, however, some caveats must be added. The optical potentials, 
{\em i.e.} the imaginary part of the self energies, are strongly energy 
dependent and
one can expect that Eq.~(\ref{eq:gshift}) is only applicable for excited
and continuum $\Lambda$-like states. 
This conclusion is based on the requirement that if all the decay channels 
of the $\Lambda$ are turned off, the system must have a stable $\Lambda$-like
ground state.

At present, the understanding of hypernuclear decay is still at a primitive
stage. Theoretical predictions for details of the dominant nonmesonic 
decay are not compatible with experimental data (see,
{\em e.g.}, Refs.~\cite{DUBACH96,PARRENO97}).
Experimental studies of relatively light hypernuclei lead to 
weakly mass-dependent lifetimes, slightly smaller than the lifetime
of a free $\Lambda$ \cite{BHANG98} in contrast to our simple estimate
in Eq.~(\ref{eq:gshift}). However, the experiments typically average over
the lowest lying states in the spectrum where we expect the lifetimes not to be
effected by the flavor mixing.
Although our discussion is an oversimplification of the problem we believe
it is useful for providing a first orientation.
A more rigorous calculation of lifetimes has to include the various decay
channels and the flavor mixing in a self-consistent manner.
This will be an important topic for future investigations.
%
%###########################################################################%
%###########################################################################%
%
\section{Summary}
\label{summary}
In this paper we study $\Lambda-\Sigma^0$ flavor mixing in hypernuclei.
Our analysis is based on a chiral effective Lagrangian containing
the baryon octet, the vector meson nonet and a light scalar singlet.
We extend the nuclear matter analysis of Ref.~\cite{MUELLER99} by
adding various tensor couplings and couplings to the electromagnetic
field which are relevant in finite systems.
The electromagnetic structure of the hyperons is described
within the theory. As a consequence, electromagnetic properties
of hypernuclei can be calculated without introducing external
form factors \cite{FST97}.

The vector coupling constants are related to the corresponding couplings
in the nonstrange sector via $SU(3)$ symmetry.
We employ a parameter set which has been obtained in a fit to properties
of normal nuclei \cite{FST97}.
The scalar hyperon couplings are determined by using
phenomenological information on energy levels in $\Lambda$ hypernuclei
and on the $\Sigma^0$ potential in nuclear matter.

The most important feature of the model is that
D-type couplings between baryons and
mesons lead to a nondiagonal self-energy in the 
$\Lambda-\Sigma^0$ sector of flavor space. As a consequence $\Lambda-\Sigma^0$ 
flavor mixing arises.

We discuss the coupled set of Dirac equations that govern the hyperons.
The solutions characterize a hypernucleus as a state of mixed flavor,
in contrast to the familiar description with distinct $\Lambda$ and 
$\Sigma^0$ particles. 
Since the mixing is relatively small, 
two qualitatively different types of solutions arise,
each dominated by a particular flavor.
However, this implies that systems which contain $\Lambda$ hyperons always have
a small admixture of $\Sigma^0$ hyperons, and vice versa.

To search for observable signatures of the effect, we study a $^{208}Pb$ 
nucleus with one strange baryon added.
Because of the large mass difference of the hyperons, only $\Lambda$-like 
states are truly bound. The $\Lambda$ part of the wave function which
describes $\Sigma$-like states is a continuum wave function.
These quasi-bound states can be found as very sharp resonances in 
scattering phase shifts.

We study the energy spectrum and electromagnetic properties of
the hyperons embedded in the nuclear medium.
The impact of the flavor mixing on the energy levels is very small,
much smaller than typical hypernuclear spin-orbit splittings.
The effect is more pronounced for electromagnetic properties.
The $\Lambda\Sigma$ transition moment leads to
deviations of hypernuclear moments from their single particle
values at the $2\%$ level.

We also examine the impact of the flavor mixing on lifetimes of
hypernuclei. Based on simple assumptions we find that the lifetimes
of excited $\Lambda$-like hypernuclei can be decreased significantly.
Although, the discussion is rudimentary it suggests that
a more rigorous examination of lifetimes, experimentally and theoretically,
could reveal signatures of $\Lambda-\Sigma^0$ mixing in strange nuclear
systems.

\acknowledgements
This work was supported in part by U.S.DOE under Grant No.
DE-FG03-93ER-40774. 
%
%###########################################################################%
%###########################################################################%
%

%
%
%###########################################################################%
%###########################################################################%
%
\begin{table}[hbt]
\caption{Hyperon coupling constants.}
\medskip
\begin{tabular}[b]{|r|l|}
$g^\omega_\Lambda$ & 7.9125  \\
$g^\varphi_\Lambda$ & 6.0381  \\
$f^\omega_\Lambda$ &  -4.7597 \\
$g^\omega_\Sigma$ &  8.2582 \\
$g^\varphi_\Sigma$ & 6.0753  \\
$f^\omega_\Sigma$ &  11.739 \\
$g^\rho_{\Lambda\Sigma}$ & 0.29942  \\
$f^\rho_{\Lambda\Sigma}$ & 14.288  \\
$\beta_D$ & -0.41754
\end{tabular}
\label{tab:coup}
\end{table}
\begin{table}[hbt]
\caption{Vector and tensor potentials.
$V^0$ and $b^0$ are the time like component of the $\omega$ and the 
$\rho^0$ mean field respectively, and $A^0$ is the Coulomb potential.}
\medskip
\begin{tabular}[b]{|r|l|}
$V_\Lambda$ & $g^\omega_\Lambda V^0 -e{\beta_D\over 6M^2}\Delta A^0$\\
$V_\Sigma$  & $g^\omega_\Sigma V^0 +e{\beta_D\over 6M^2}\Delta A^0$\\
$V_{\Lambda\Sigma}$ & $g^\rho_{\Sigma\Lambda} b^0 
+ e{\beta_D\over 2\sqrt{3}M^2}\Delta A^0$\\
\hline
$T_\Lambda$ & ${1\over 2M}
(f^\omega_\Lambda V^0{}^\prime +e\mu_\Lambda A^0{}^\prime)$\\
$T_\Sigma$  & ${1\over 2M}
(f^\omega_\Sigma V^0{}^\prime +e\mu_{\Sigma^0} A^0{}^\prime)$\\
$T_{\Lambda\Sigma}$  & ${1\over 2M}
(f^\rho_{\Lambda\Sigma} b^0{}^\prime +e\mu_{\Lambda\Sigma^0} A^0{}^\prime)$
\end{tabular}
\label{tab:pots}
\end{table}
%
%###########################################################################%
%###########################################################################%
%
\section*{Figures}
\global\firstfigfalse
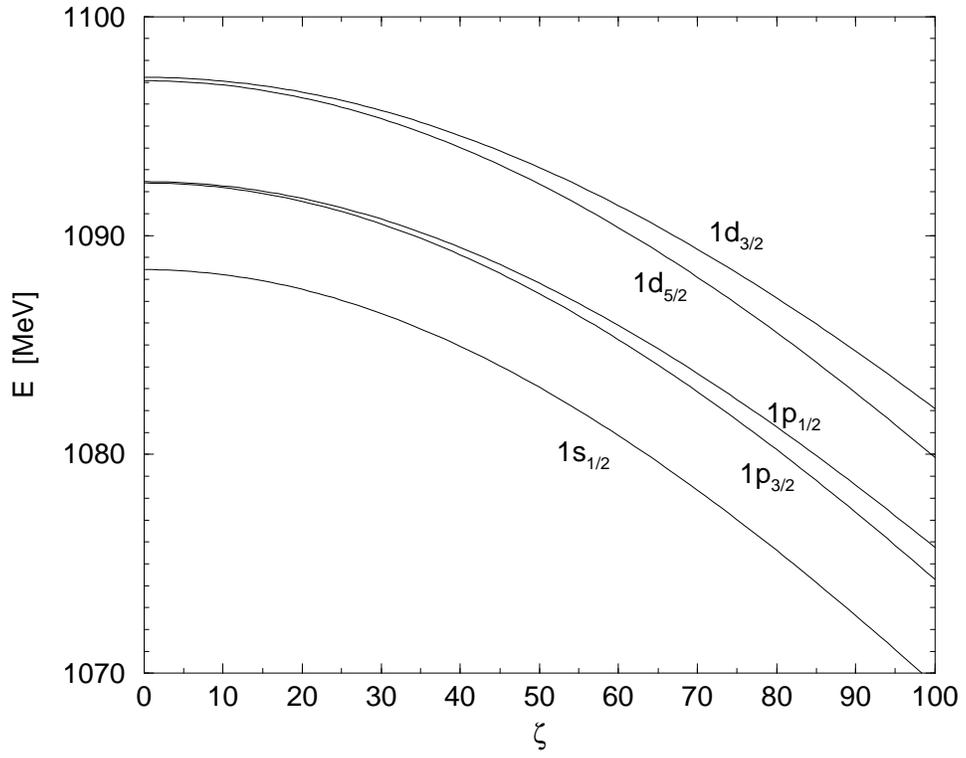
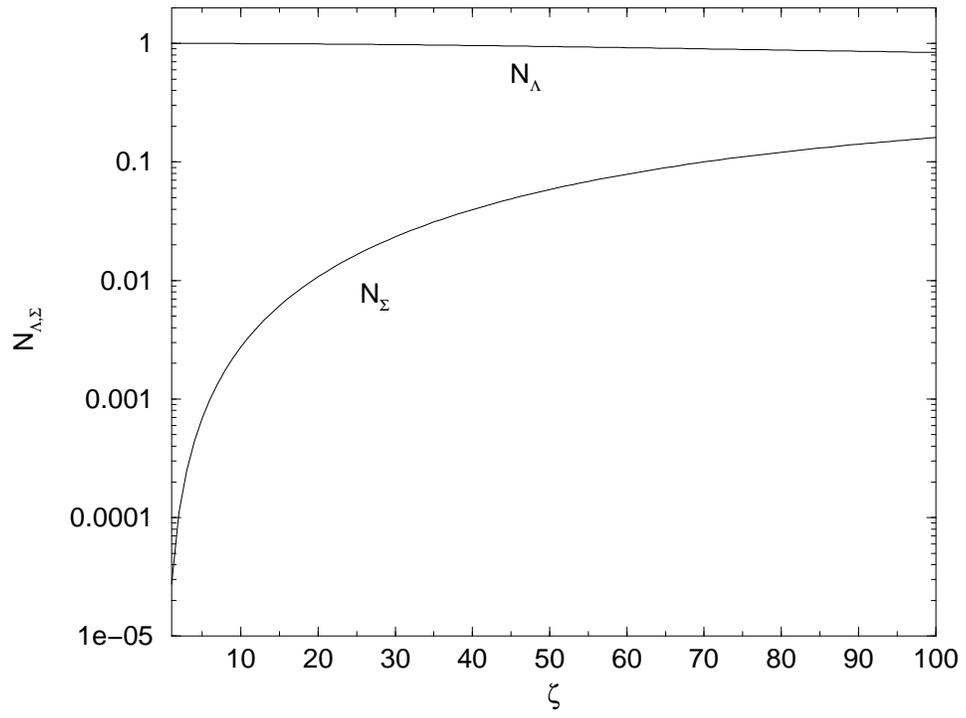
\begin{figure}[tbhp]
\caption{(a) Binding energies of several $\Lambda$ like states. 
The mixing potentials are scaled by $\zeta$.
(b) Flavor fractions of the $1s_{1/2}$ state.}
\label{fig:spec}
\end{figure}
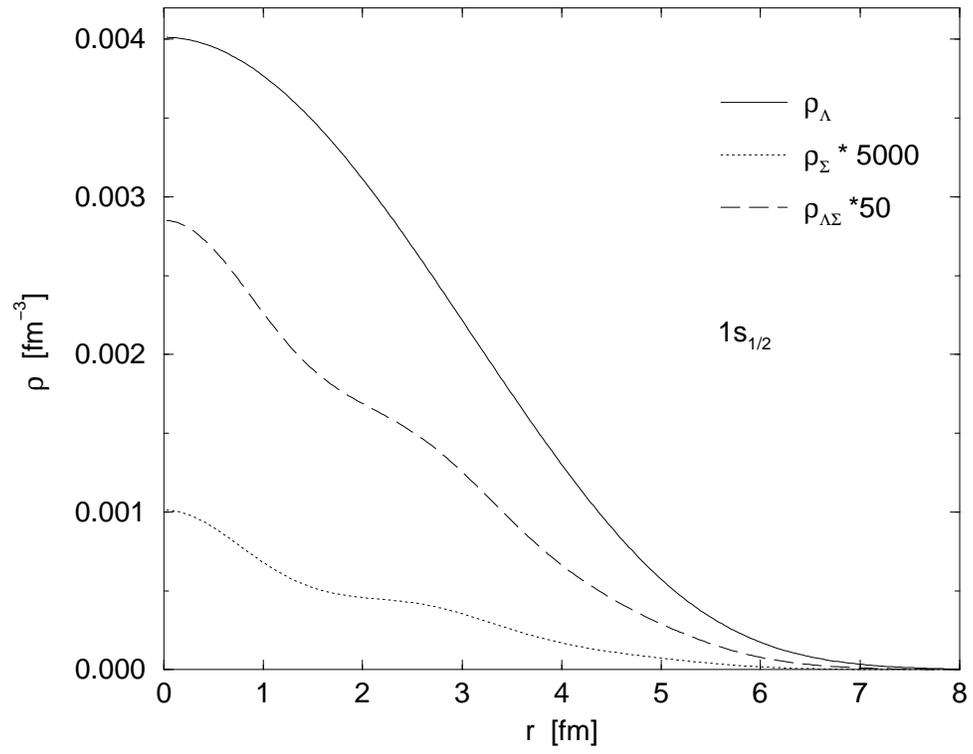
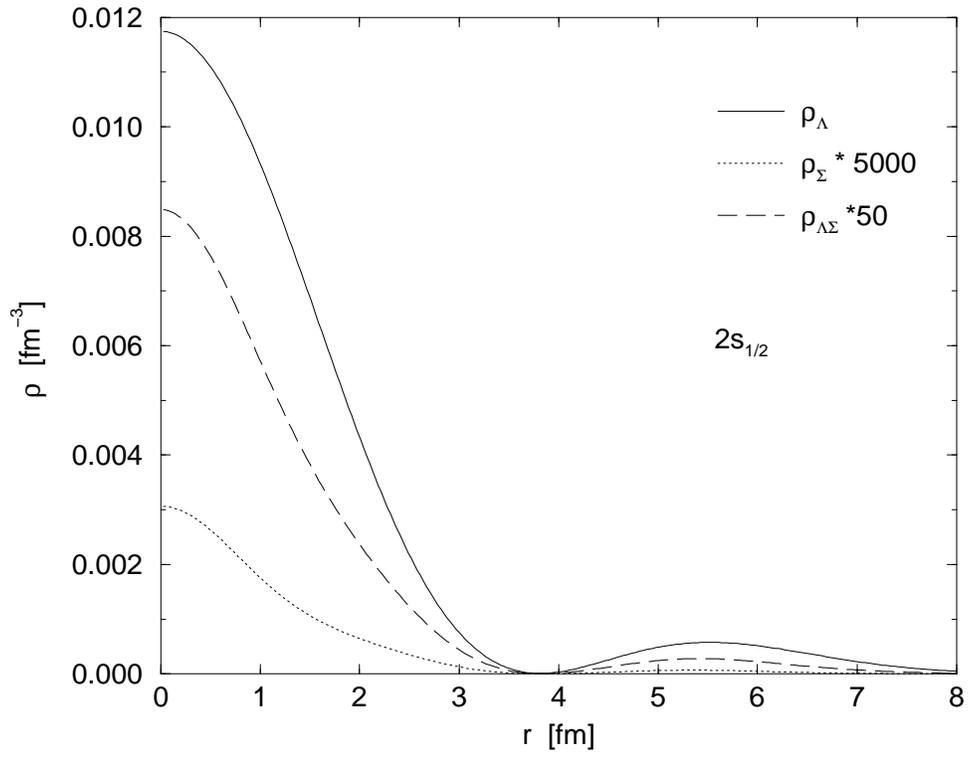
\begin{figure}[tbhp]
\caption{Pure flavor and mixed flavor baryon densities. (a) shows a $1s_{1/2}$ 
state and (b) a $2s_{1/2}$ state.} 
\label{fig:rho}
\end{figure}
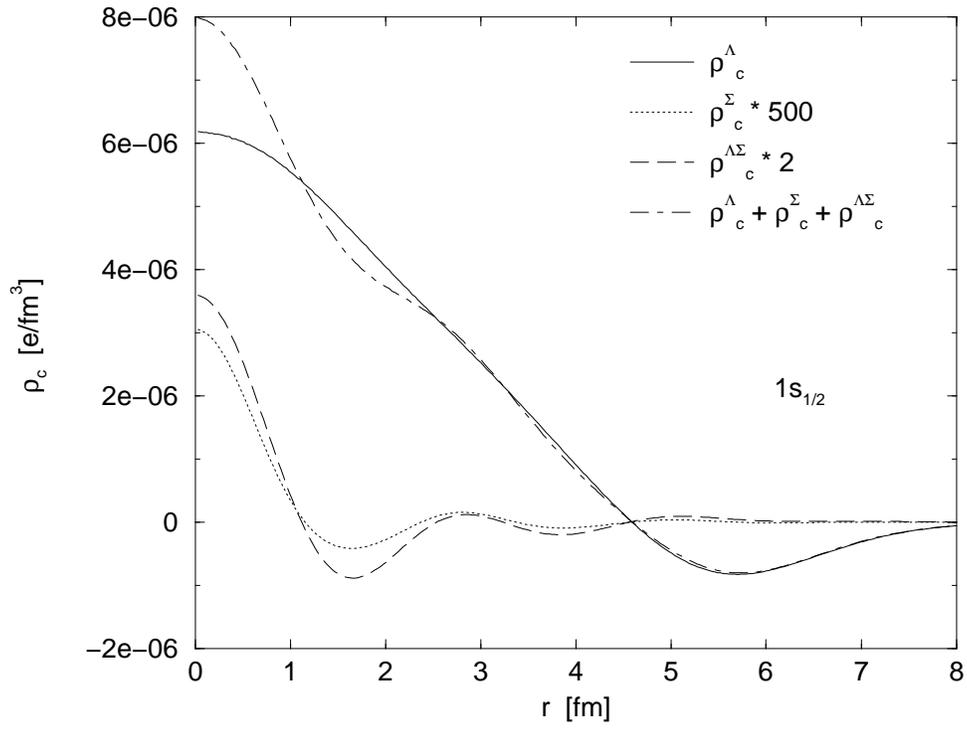
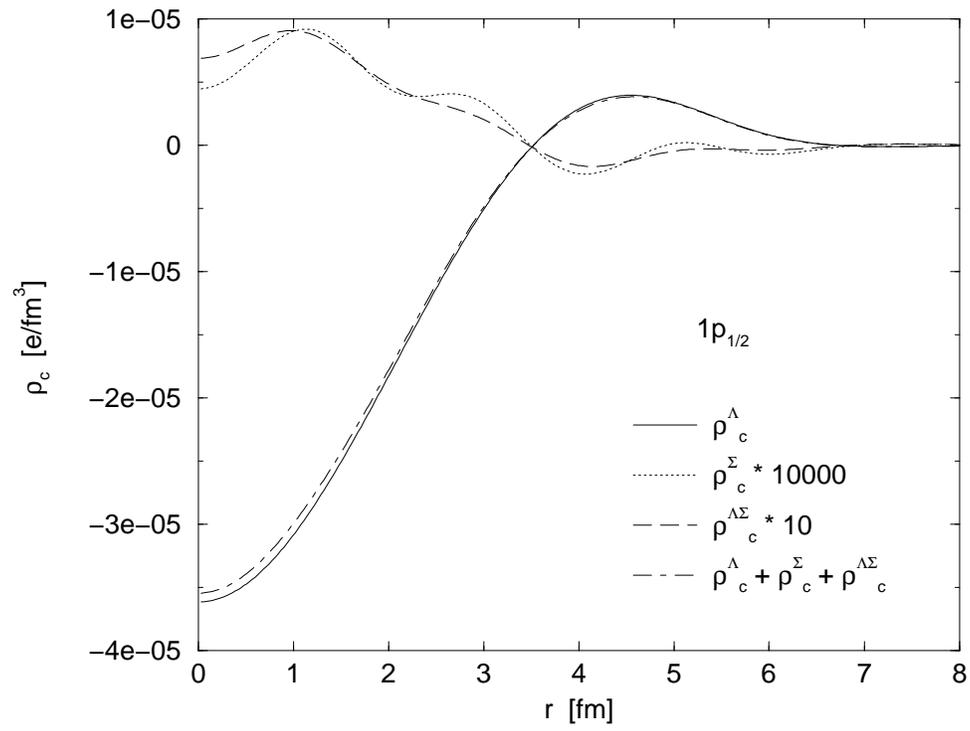
\begin{figure}[tbhp]
\caption{$\Lambda$, $\Sigma$ and mixed contribution to the charge density.
(a) shows a $1s_{1/2}$ state and (b) a $1p_{1/2}$ state.}
\label{fig:rhoc}
\end{figure}
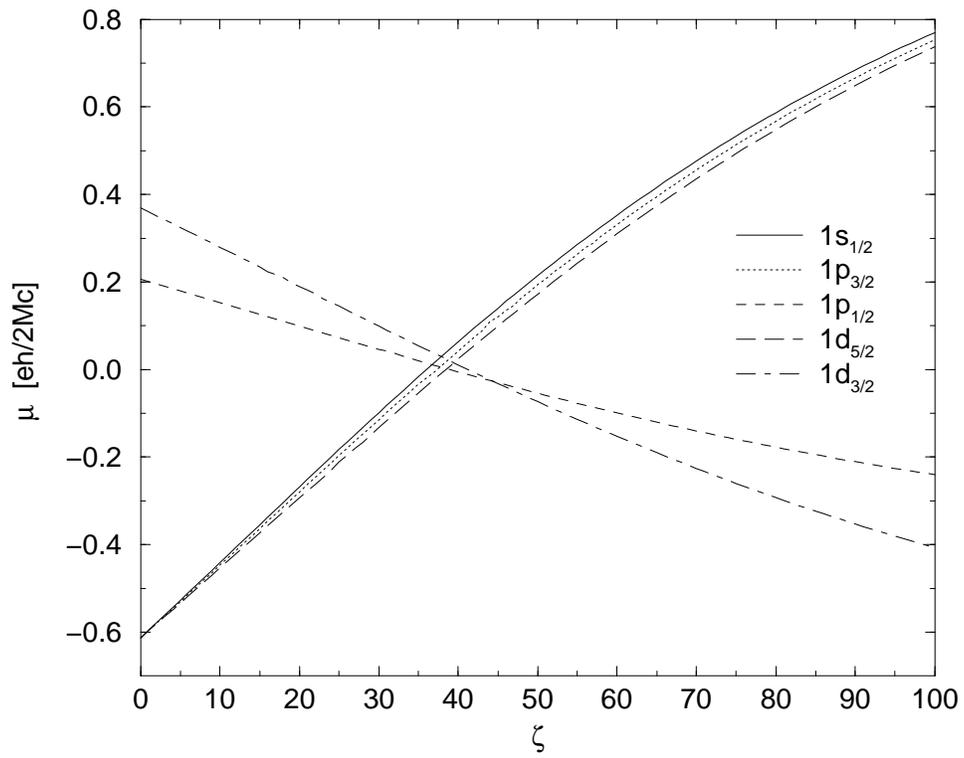
\begin{figure}[tbhp]
\caption{Magnetic moments for several $\Lambda$ like states. 
The mixing potentials are scaled by $\zeta$.}
\label{fig:mag}
\end{figure}
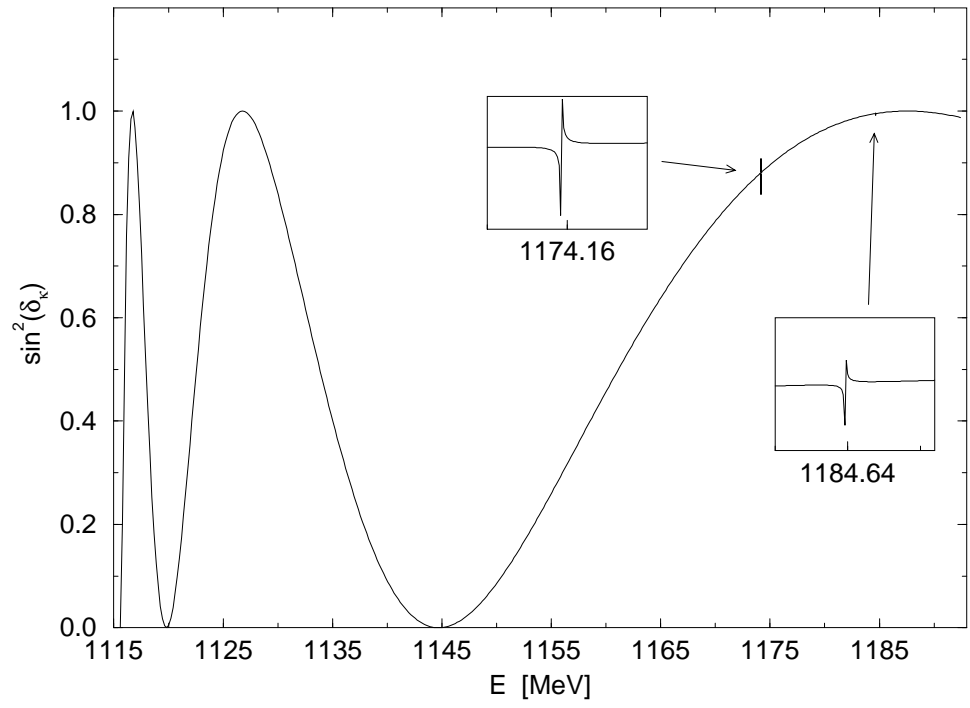
\begin{figure}[tbhp]
\caption{The function $\sin^2(\delta_\kappa)$ for a $1p_{3/2}$ state.
The resonances at 1174.16 MeV and 1184.64 MeV are signatures of pure
$\Sigma^0$ bound states which arise when the mixing is turned off.}
\label{fig:sin2}
\end{figure}
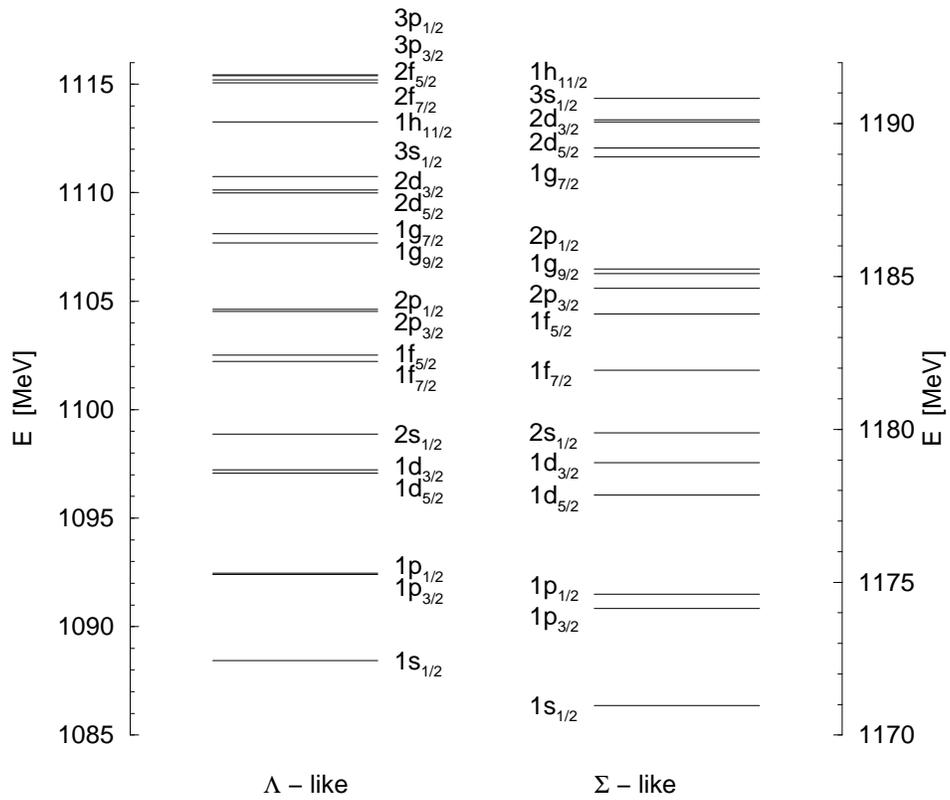
\begin{figure}[tbhp]
\caption{$\Lambda$-like and $\Sigma$-like spectrum for the $^{208}Pb$
core nucleus.}
\label{fig:spec2}
\end{figure}

\begin{references}
%
\bibitem{BANDO90} H. Band\={o}, T. Motoba and J. \u{Z}ofka,
                  {\em Int. J. of Mod. Phys. A {\rm 5}} (1990) 4021.
%
\bibitem{RAYET81} M. Rayet, Nucl. \ Phys.\ A 367 (1981) 381.
%
\bibitem {RUFA87} M. Rufa,  H. St\"ocker, J. A. Maruhn, W. Greiner,
                  and P. G. Reinhard, 
                  J. \ Phys. G 13 (1987) L143.
%
\bibitem {MARES89} J. Mare\v{s} and J. \v{Z}ofka, Z.\ Phys.\ A 333 (1989) 209.
%
\bibitem{SCHAFFNER93} J. Schaffner, C. B. Dover, A. Gal, C. Greiner,
                      and H. St\"ocker,
                      Phys.\ Rev.\ Lett. \ 71 (1993) 1328.
%
\bibitem{PAPAZOGLOU98a} P. Papazoglou, S. Schramm, J. Schaffner-Bielich, 
                       H.St\"ocker and W. Greiner, 
                       Phys.\ Rev.\ C 57 (1998) 2576.
%
\bibitem{PAPAZOGLOU98b}P. Papazoglou, D. Zschiesche, S. Schramm, 
                       J. Schaffner-Bielich, H.St\"ocker and W. Greiner, 
                       Phys.\ Rev.\ C 59 (1999) 411.
%
\bibitem {MUELLER99} H. M\"uller, Phys.\ Rev.\
                     C 59 (1999) 1405.
%
\bibitem {FST97} R. J. Furnstahl, B. D. Serot, and H.-B. Tang, Nucl.\ Phys.\
                   A 615  (1997) 441.
%
\bibitem{SEROT97} B. D. Serot and J. D. Walecka, 
                  {\em Int. J. of Mod. Phys. E {\rm 6}} (1997) 515.
%
\bibitem{BILENKY78} S. M. Bilenkii and B. Pontecorvo, Phys.\ Rep.\ 41
                    (1978) 225. 
%
\bibitem{COLEMAN61} S. Coleman and S. L. Glashow, Phys. \ Rev. \ Lett. \ 6
                  (1961) 423.
%
\bibitem{PDG} C. Caso {\em et al.}, The \ European \ Physical \ Journal \ C 3 
              1 (1998).
%
\bibitem{MEISSNER97} U.-G. Mei\ss{}ner and S. Steininger,
                     Nucl.\ Phys.\ B 499 (1997) 349.
%
\bibitem{CHRIEN88} R. E. Chrien, Nucl. \ Phys. \ A 478 (1988) 705c.
%
\bibitem{SCHAFFNER94} J. Schaffner, C. B. Dover, A. Gal, C. Greiner,
                      D. J. Millener and H. St\"ocker,
                      Ann.\ Phys.\ (N.Y.) 235 (1994) 35.
%
\bibitem{MSW}  L. Wolfenstein, Phys.\ Rev.\ D 17 (1978) 2369.\\
               S. P. Mikheev and A. Yu. Smirnov, Sov.\ J.\ Nucl.\ Phys. \
               42 (1985) 913.
%
\bibitem{DOVER95} C. B. Dover, H. Feshbach and A. Gal,
                  Phys.\ Rev.\ C 51 (1995) 541.
%
\bibitem{MARES94} J. Mare\v{s} and B. K. Jennings, Phys. \ Rev. C 49 (1994)
                  2472.
%
\bibitem{GATTONE91} A. O. Gattone, M. Chiapparini and E. D. Izquierdo,
                    Phys. \ Rev. C 44 (1991) 548.
%
\bibitem{COHEN92} J. Cohen and J. V. Noble, Phys. \ Rev. C 46 (1992) 801.
%
\bibitem{SHEPARD88} J. R. Shepard, E. Rost, C.-Y. Cheung and J. A. McNeil,
                    Phys. \ Rev. C 37, 1130 (1988).
%
\bibitem{NAKAI96} K. Nakai, JHP-Supplement-22 (1996) 31.
% 
\bibitem{DOVER89} C. B. Dover, D. J. Millener and A. Gal, Phys. \ Rep. \ 184
                  (1989) 1.
%
\bibitem{ADELBERGER77} E. G. Adelberger {\em et al}, Phys. \ Rev. C 15
                       (1977) 484.
%
\bibitem{FLANZ79} J. B. Flanz {\em et al}, Phys. \ Rev. \ Lett. 43
                  (1979) 1922.
%
\bibitem{DUBACH96} J. F. Dubach, G. B. Feldman, B. R. Holstein
                   and L. De La Torre, Ann.\ Phys.\ (N.Y.) 249 (1996) 146.
%
\bibitem{PARRENO97} A. Parren\~{n}o, A. Ramos and C. Bennhold,
                    Phys. \ Rev. \ C 56 (1997) 339.
%
\bibitem{BHANG98} H. Bhang {\em et al}., Phys. \ Rev. \ Lett. \ 81
                  (1998) 4321.
\end{references}
\end{document}